\documentclass[preprint,12pt,sort&compress]{elsarticle}
\usepackage[a4paper, left=2cm, right=2cm]{geometry}

\usepackage{amssymb}
\usepackage{amsmath}
\usepackage{comment}
\usepackage{picture}
\usepackage{makecell}
\usepackage{booktabs}

\let\oldAA\AA
\renewcommand{\AA}{\text{\oldAA}}

\usepackage{xspace}
\usepackage{multicol}
\usepackage{xcolor}
\usepackage[normalem]{ulem}
\usepackage[flushleft]{threeparttable}
\usepackage{gensymb}
\usepackage{textcomp,mathcomp}
\usepackage{url}
\usepackage{todonotes}
\usepackage{hyperref}
\newcommand{\etal}{\emph{et al.}\xspace}

\newcommand{\stkout}[1]{\ifmmode\text{\sout{\ensuremath{#1}}}\else\sout{#1}\fi}

\journal{Scripta Materialia}

\begin{document}

\begin{frontmatter}

\title{Interstitials as a key ingredient for P segregation to grain boundaries in polycrystalline $\alpha$-Fe}

\author[inst1]{Amin Reiners-Sakic}

\affiliation[inst1]{organization={Department of Materials Science, Montanuniversität Leoben},
            addressline={Franz Josef-Straße 18}, 
            city={8700 Leoben}, 
            country={Austria}}
\affiliation[inst2]{organization={Christian Doppler Laboratory for Knowledge-based Design of Advanced Steels, Department of Materials Science, Montanuniversität Leoben},
            addressline={Franz Josef-Straße 18}, 
            city={8700 Leoben}, 
            country={Austria}}
\author[inst1]{Alexander Reichmann}
\author[inst1]{Christoph Dösinger}
\author[inst1]{Lorenz Romaner}
\author[inst2]{David Holec}

\begin{abstract}
Solute segregation to grain boundaries (GBs) significantly impacts material behavior, with most studies focusing on substitutional solute segregation while neglecting interstitial segregation due to its increased complexity. The site preference, interstitial or substitutional, for P segregation in $\alpha$-Fe still remains under debate. This work investigates both substitutional and interstitial GB segregation in a polycrystalline model using classical interatomic potentials and machine learning. The method is validated with H and Ni, whose segregation behaviors are well understood. For P, we find segregation to both GB site types, with a preference for substitutional sites based on mean segregation energy. However, the abundance of interstitial sites means interstitial segregation also significantly contributes to the GB enrichment with P. This highlights the importance of considering interstitial P segregation alongside substitutional segregation. Additionally, obtaining a representative spectrum of segregation energies is crucial for accurate, experimentally aligned predictions.
\end{abstract}

\begin{graphicalabstract}
\includegraphics[width=\textwidth]{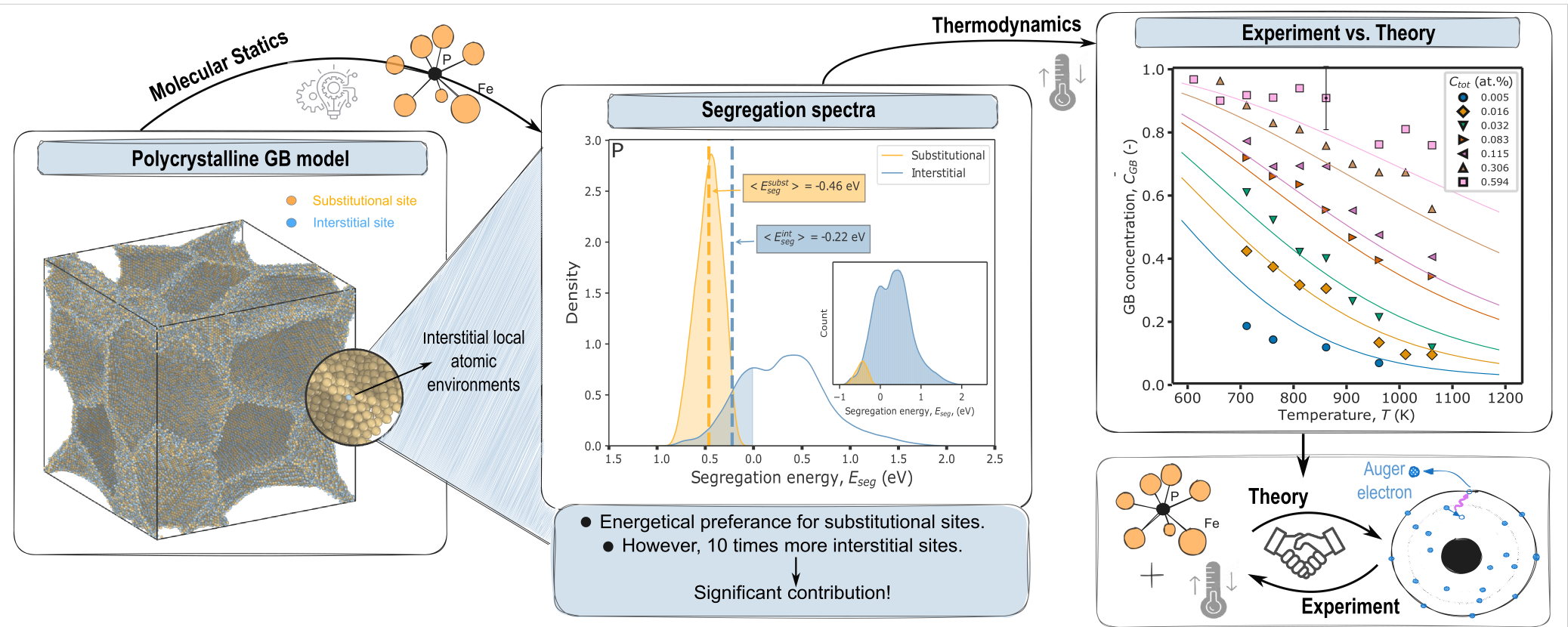}
\end{graphicalabstract}

\begin{highlights}
\item Comprehensive study of substitutional and interstitial segregation profiles in an atomistic polycrystalline grain boundary model of $\alpha$-Fe.
\item Segregation of P, measured by mean segregation energy, to substitutional sites is stronger than to interstitials.
\item Interstitial sites nevertheless significantly contribute to the GB segregation of P due to their large number as compared with substitutional sites.
\end{highlights}
\end{frontmatter}

Segregation of solute atoms and impurities can significantly affect material properties, for example, by strengthening or weakening the grain boundary (GB) network or altering the corrosion resistance. This has led to extensive research on solute segregation using both experimental methods—such as Auger Electron Spectroscopy (AES), Atom Probe Tomography (APT), and Energy-Dispersive X-ray Spectroscopy (EDX)—and computational approaches, most notably Density Functional Theory (DFT) and Molecular Dynamics (MD) simulations. 
In some systems, a good agreement has been achieved between the experimentally determined and simulated GB excess~\cite{scheiber2015ab}. However, in other systems, strong deviations between experimental and computational results are observed~\cite{reichmann2024grain}. A key factor contributing to these differences is the resolution and interpretation of segregation energies.
In experiments, the resolution typically spans from nanometers to micrometers, resulting in the determination of an averaged or effective segregation energy derived from concentration profiles of solute atoms at GBs. This profile may correspond to a single GB, as is often the case in APT measurements~\cite{ebner2021grain}, or it represents an average over several GBs, as in AES measurements of fractured surfaces~\cite{grabke1987effects}. In contrast, computational methods like DFT can provide a set of segregation energies~\cite{scheiber2015ab, scheiber2021impact, sakic2024interplay, mai2022segregation, mai2023phosphorus}, as they allow solutes to be placed at distinct, atomistically resolved sites along the GB. This capability reveals another critical distinction; while experimental techniques typically lack the resolution to differentiate between segregation at substitutional or interstitial sites at GBs, computational approaches can, and for atomistic methods, in fact, must resolve the fundamental difference between these two types of segregation sites. 

One alloy system where the site preference of solutes at GBs remains a topic of ongoing discussion despite numerous investigations is Fe-P. On the one hand, \emph{ab initio} calculations indicate a higher tendency for substitutional segregation of P at special coincident site lattice GBs~\cite{mai2022segregation, scheiber2021impact, yamaguchi2007decohesion}. On the other hand, Lejček~\etal~\cite{lejvcek2021entropy, lejvcek2016interstitial}, who extracted enthalpy and entropy of segregation for P in $\alpha$-Fe based on experimentally measured GB concentrations using AES, found that these values more closely align with those of typical interstitial solutes like C, rather than typical substitutional segregants like Si or Al. Based on the enthalpy-entropy compensation effect, they further concluded that P undergoes an entropy-driven transition from substitutional to interstitial site preference at GBs above 700\,K~\cite{lejvcek2016interstitial}. Scheiber~\etal~\cite{scheiber2021impact} provided further insights by considering P segregation at four different special coincident site lattice GBs. Their study demonstrated that the strong temperature dependence of experimentally determined segregation energies is, to a large part, a result of applying the single-site McLean isotherm~\cite{mclean1958grain}, which neglects the multi-site nature of segregation. By averaging over the different segregation sites, an artificial temperature dependence emerges. Recently, Reichmann~\etal~\cite{reichmann2024grain} used Bayesian inference to derive a segregation energy spectrum for P in $\alpha$-Fe based on AES-measured concentration profiles from Erhart and Grabke~\cite{erhart1981equilibrium}. They also compared the experimentally measured GB concentrations with those derived from atomistic simulations ~\cite{scheiber2021impact}, including the spectral nature of segregation energies, segregation vibrational entropy, and P-P interactions for substitutional segregation sites. The comparison revealed a gap between the experimental and computational results, raising new questions about the nature of P segregation at GBs.

In this work, we aim to resolve the ongoing debate regarding the site preference of P at GBs in $\alpha$-Fe by investigating the segregation spectrum for both interstitial and substitutional P in a polycrystalline model. First, we seek to clarify the site preference of P at GBs. Second, we make use of the complete segregation profile—including both interstitial and substitutional segregation energies—to calculate GB concentrations at different temperatures and bulk concentrations. These results are then compared with the experimental findings of Erhart and Grabke~\cite{erhart1981equilibrium} and the computational results of Reichmann~\etal~\cite{reichmann2024grain}. To achieve this, we utilize recent advances in machine learning techniques for predicting segregation energy spectra~\cite{wagih2020learning} and extend these methods to predict interstitial segregation energies in body-centered cubic materials. We also predict the segregation profiles for H and Ni. For these solutes, the segregating behavior is well established in the literature, providing an opportunity to validate the approach.
\newline\newline
The polycrystal of size $20\times20\times20\,$nm was constructed with the help of \texttt{Atomsk}~\cite{hirel2015atomsk} using 12 initial seeds for the location and orientation of the grains, yielding a structure consisting of $\text{699}\,\text{323}$ atoms. First, the polycrystal was relaxed using the conjugate-gradient algorithm and then isothermally annealed for 500$\,$ps at 600$\,$K. This was followed by quenching to 0$\,$K with a cooling rate of $3 \text{ K/ps}$ and a final relaxation with the force criterion set to $10^{-8} \text{ eV/\AA}$. This procedure is inspired by previous works~\cite{reichmann2024grain, wagih2020learning}. These calculations were performed with the \texttt{LAMMPS} package~\cite{LAMMPS} and Mendelev’s embedded-atom-method (EAM) potential for the Fe-Fe interactions~\cite{mendelev2003development}.

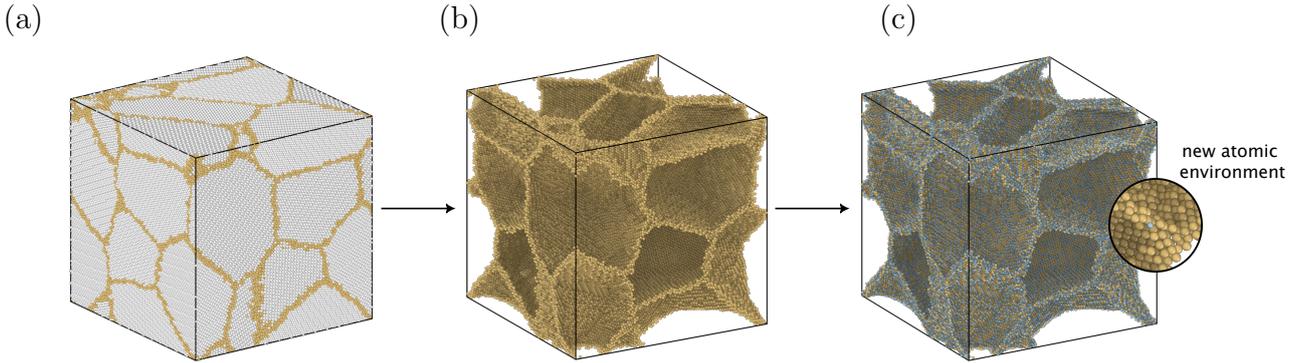
\begin{figure}[!ht]
    \centering
    \begin{picture}(0,0)
        \put(-238,0){(a)}
        \put(-75,0){(b)}
        \put(90,0){(c)}
    \end{picture}\smallskip\\
    \includegraphics[width=1\textwidth]{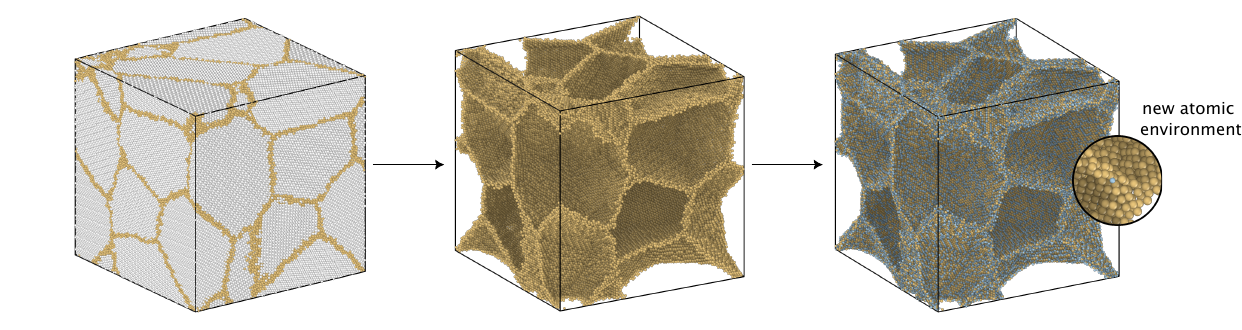}
    \caption{Workflow for the generation of interstitial GB sites, starting from a) the polycrystalline model, where the orange atoms indicate the GB atoms obtained from the CNA; b) a model showing only the GB atoms used in the Voronoi analysis to identify the interstitial sites; and c) the polycrystalline model showing the GB network for both substitutional (orange) and interstitial segregation (blue). All visualizations were done using \texttt{ovito}~\cite{ovito}.}
    \label{fig:methodology}
\end{figure}

In order to obtain the interstitial sites, we performed three consecutive steps (see Fig.~\ref{fig:methodology}). First, for the selection of the GB atoms, we utilized the adaptive common neighbor analysis (CNA)~\cite{stukowski2012structure, ovito}. This resulted in $\text{108}\,\text{053}$ GB atoms, which were then used for the characterization of the substitutional segregation spectrum. The GB atoms are highlighted in orange in Fig.~\ref{fig:methodology}a. In the next step, we removed the remaining bulk atoms from the polycrystal, leaving only the GB atoms (Fig.~\ref{fig:methodology}b) and performed Voronoi tessellations (using the \texttt{scipy} implementation) to identify the interstitial sites. To do so, we followed the same principles as in the perfect bcc Fe, where the tetrahedral sites are given by the vertices points of the Voronoi polyhedral, and the octahedral sites are given by the midpoint of an atom and its next-nearest neighbor atom. To reduce the number of interstitial sites, we used spatial analysis to remove interstitial positions that are further away from a neighboring host atom than the relaxed bulk Fe lattice parameter of 2.839$\,\AA$. This resulted in a total of $1\,191\,555$ interstitial sites.

The segregation energy spectra were computed using a recently proposed machine learning approach~\cite{wagih2020learning}, which utilizes smooth overlap of atomic positions (SOAP)~\cite{bartok2013representing} features to describe the local atomic environments (LAEs) at the GB. Spherical Gaussian-type orbitals were employed as radial basis functions with $n_{\text{max}} = 12$, $l_{\text{max}} = 9$, and a cutoff radius ($r_{\text{cut}}$) of 6\,\AA{}, yielding 780 features per atom. These calculations were performed using the \texttt{DScribe} library~\cite{dscribe2}. To construct the SOAP feature vector for the interstitial sites, each interstitial site was introduced separately into the polycrystal thereby neglecting the presence of neighboring interstitials. This approach allowed us to isolate and characterize the LAEs of each interstitial site independently. The dimensionality of the obtained SOAP feature vectors was then reduced to ten principal components using Principal Component Analysis (PCA). By means of \textit{k}-mean clustering, 100 representative substitutional and 300 representative interstitial sites were selected for which the segregation energies were computed using molecular statics applying EAM potentials, Fe-P~\cite{ackland2004development}, Fe-H~\cite{wen2021new}, and Fe-Ni~\cite{bonny2009fe}.
The calculated segregation energies were then used to predict the remaining substitutional and interstitial segregation energies using random forest regression. All machine learning methods (PCA, \textit{k}-means, and random forest regression) were implemented using the \texttt{scikit-learn} library~\cite{scikit-learn}.

\begin{figure}[!ht]
    \centering
    \includegraphics[width=1\linewidth]{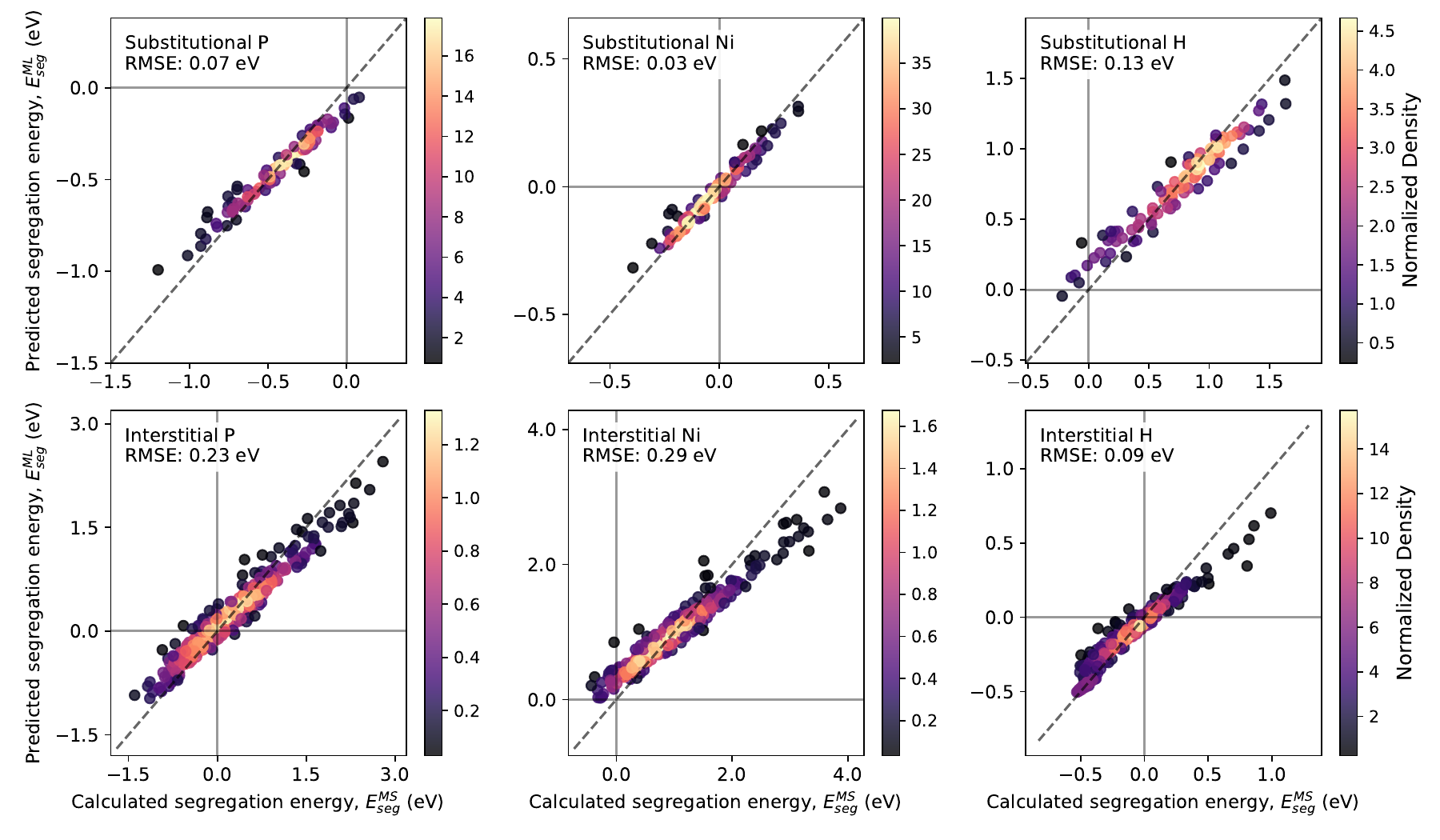}
    \caption{Performance of the ML model for segregation energy predictions. The shown points and the respective RMSE values are based on the training data set. The predicted segregation energies are plotted against the calculated segregation energies for P (left column), Ni (middle), and H (right) in the substitutional (top row) and interstitial (bottom row) GB sites. The dashed line indicates a perfect fit.}
    \label{fig:calc_vs_pred}
\end{figure}

We note that the segregation energies were calculated only for transitions from the most stable bulk site (see Supplementary Material Fig.~S1 for the overview of solute formation energies), i.e., tetrahedral for H and substitutional for P and Ni, to the GB site of interest, which could be either substitutional or interstitial. Hence, the calculation of the segregation energy $\Delta E_{seg}^k(X)$ of solute $X$ to site $k$ can be summarized as follows:
\begin{equation}
    \Delta E_{seg}^k(X) = E_{poly}^k[\text{Fe} + X] + E_{bulk}[\text{Fe}] + \delta\frac{E_{bulk}[\text{Fe}]}{m} - E_{poly}[\text{Fe}] - E_{bulk}[\text{Fe} + X],
    \label{eqn:equation_segregation}
\end{equation}
where $E_{poly}^k[\text{Fe} + X]$, and $E_{poly}[\text{Fe}]$ are the total energies of the polycrystal with and without a solute, respectively. Similarly, $E_{bulk}[\text{Fe} + X]$ and $E_{bulk}[\text{Fe}]$  are the total energies of the bulk bcc-Fe cells with a solute (either at an interstitial or substitutional site) and without a solute, respectively. The fractional term accounts for different segregation scenarios, where $m$ is the total number of bulk atoms.
The factor $\delta$ of the per-atom energy of Fe ($E_{bulk}[\text{Fe}]/m$) depends on the specific transition and is there to account for possible changes in the number of atoms. The transition refers to the change of the solute position (interstitial or substitutional) when segregating from bulk to the GB. The four possible scenarios are summarized in Table~\ref{tab:Segregation_scenarios}. 
\begin{table}[!ht]
    \centering
    \caption{Dependence of the factor $\delta$ from Eq.~\ref{eqn:equation_segregation} on the segregation path.}
    \begin{tabular}{c c c}
    \hline
    Bulk site & GB site & $\delta$ \\
    \hline
    Substitutional & Substitutional & 0\\
    Substitutional & Interstitial & $-1$ \\
    Interstitial & Substitutional & 1 \\
    Interstitial & Interstitial & 0  \\   
    \hline
    \end{tabular}
    \label{tab:Segregation_scenarios}
\end{table}
\newline
Based on the obtained site-resolved segregation energies $\Delta E_{seg}^k(X)$, we can apply an extension of the McLean equation~\cite{mclean1958grain}, as formulated by White and Coghlan~\cite{white1977spectrum}, to calculate the site-specific solute concentration at the GB, $C_{GB}^k(X)$, expressed as:
\begin{equation}
\label{eqn:multi-siteMCLean}
   \frac{C_{GB}^k(X)}{1 - C_{GB}^k(X)} = \frac{C_{bulk}(X)}{1 - C_{bulk}(X)}\exp\left(-\frac{\Delta E_{seg}^k(X)}{k_BT}\right)
\end{equation}
where $C_{bulk}(X)$ indicates the solute concentration in the bulk, $k_B$ is the Boltzmann constant, and $T$ is the temperature. Further, $C_{bulk}(X)$ can be substituted by the global composition $C_{tot}(X)$, given by 
\begin{equation}
C_{tot}(X) = (1-f_{GB})C_{bulk}(X)+f_{GB}C_{GB}(X)
\end{equation}
where the fraction of GB sites is given by the factor $f_{GB}$, which can reasonably be assumed to be $10^{-6}$ for large grain diameters~\cite{reichmann2024grain}. 
To obtain an average GB solute concentration $\bar{C}_{GB}(X)$ of solute $X$, the total amount of segregated P atoms is calculated from a summation over $C_{GB}^k(X)$ of all sites (interstitial and substitutional combined) and divided by the total number of atoms in the GB (Fe and P) $N_{GB}$\cite{scheiber2021impact}:
\begin{equation}
\label{eqn:Conc_multisite}
 \bar{C}_{GB}(X)  = \frac{\sum_{k} C^k_{GB}}{N_{GB}}.
\end{equation}
Through the temperature dependence of the site-specific solute concentration, the average GB concentration changes with temperature. Note that $N_{GB}=N_{GB}^{subst}+N_{GB}^{int}$. While $N_{GB}^{subst}$ corresponds to the amount of substitutional sites in the GB, $N_{GB}^{int}$ is not equal to the total amount of interstitial sites but equal to $N_{GB}^{int}=\sum_{k\in int}C^k_{GB}(X)$ since the interstitial site remains empty if not occupied by the solute. This accounts for the fact that unoccupied interstitial sites do not contribute to the total concentration~\cite{scheiber2021impact}. 
\newline\newline
The performance of the machine learning models in predicting both substitutional and interstitial segregation energies, based on the training data set ($\equiv$ clustering points) is shown in Fig.~\ref{fig:calc_vs_pred}. A clear trend in accuracy can be observed. For substitutional solute segregation in Fe, Ni exhibits the highest accuracy, followed by P and H, whereas for interstitial solute segregation, the order is reversed. This suggests that the greater the size mismatch between the GB site size and the atomic radius of the solute (i.e., the more structural relaxation taking place), the higher the uncertainty in the model’s predictions.

For example, the mean Voronoi radius for the substitutional GB sites (evaluated on our model) is $1.42\,\AA$. The atomic radii of Fe ($1.40\,\AA$) and Ni ($1.35\,\AA$) are very similar to this value, whereas the atomic radius of P ($1.00\,\AA$) and H ($0.25\,\AA$) are significantly smaller. This indicates that the greater the size mismatch, the more structural relaxation happens when the solute is placed at a substitutional lattice site. 
On the contrary, the interstitial radius of tetrahedral and octahedral sites on $\alpha$-Fe are $0.36\,\AA$ and $0.19\AA$, respectively.  Here, H fits best, followed by P and Ni. Therefore, interstitial Ni is expected to induce the largest amount of structural relaxation compared to the others.

\begin{figure}[!ht]
    \centering
    \includegraphics[width=8cm]{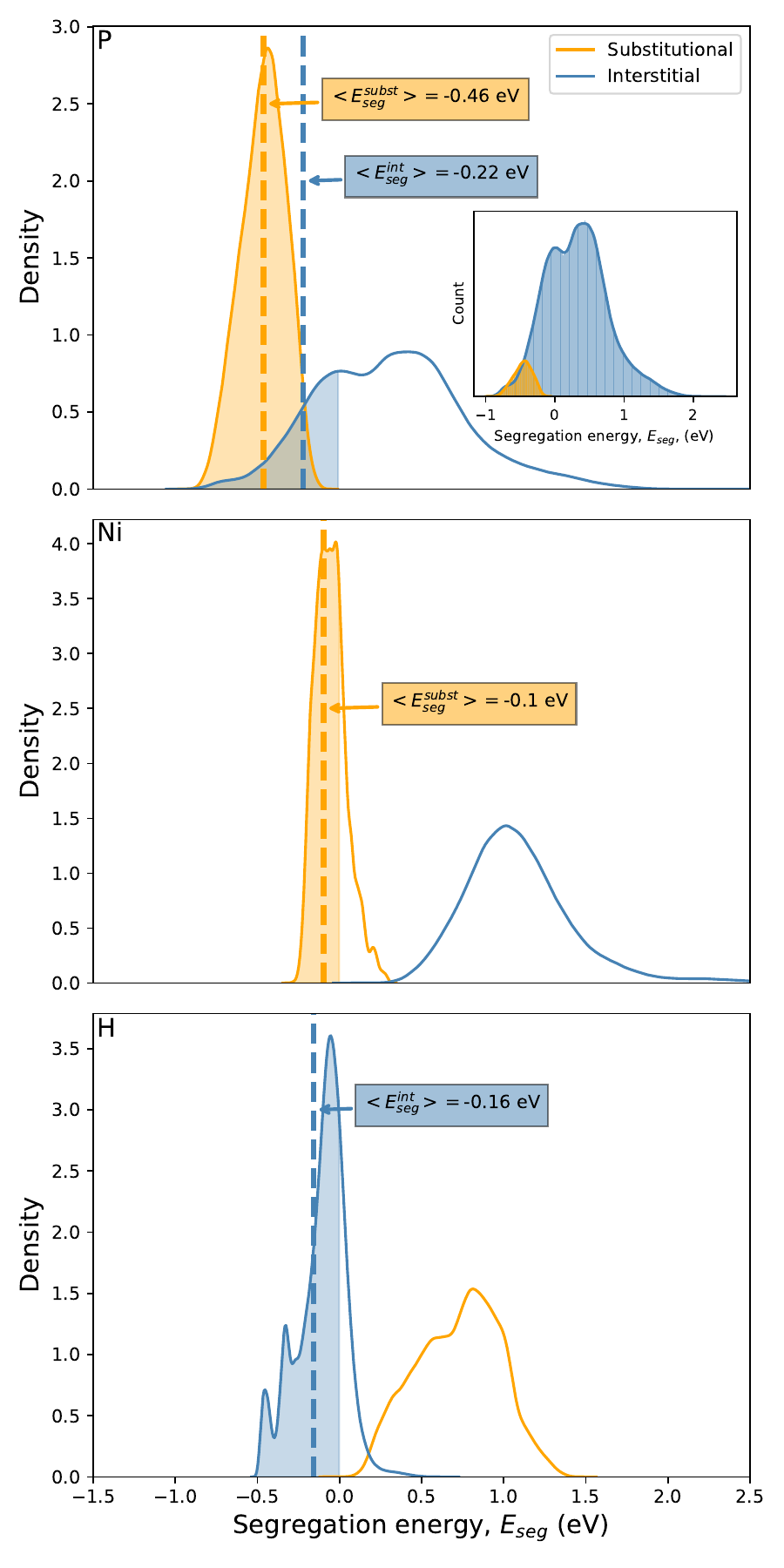}
    \caption{Normalized segregation spectra for the elements P, Ni, and H. The orange and blue distributions represent the substitutional and interstitial segregation spectra, respectively. The dashed lines indicate the mean negative segregation energies. The inset plot for P shows that the actual number of interstitial sites with similar energies to the substitutional segregation sites is significantly higher, illustrating the dominance of interstitial segregation in these regions.}
    \label{fig:segregation_profiles}
\end{figure}

A similar trend has been reported in previous studies on substitutional segregation in other solvent-solute systems~\cite{huber2018machine}. This behavior arises because the calculated segregation energy is being fitted to a set of descriptors derived from the undecorated GB model. However, as the size mismatch increases, nonlinear contributions to the elastic energy during relaxation become more significant. Consequently, the model's accuracy decreases not only when the solute is much larger than the GB site---such as Pb segregating to substitutional Al GB sites~\cite{huber2018machine} or, in our case, interstitial Ni segregation---but also when smaller atoms fill GB sites with large volume, as observed in the case of substitutional H segregation.
These findings highlight the need for additional descriptors capable of capturing large atomic relaxations to improve the predictive accuracy of ML models for such systems. However, addressing this challenge is beyond the scope of this work and should instead serve as motivation for future research in this field. For now, we will focus on the obtained segregation spectra for the three elements P, Ni, and H, as shown in Fig.~\ref{fig:segregation_profiles}. 

In the case of P, the segregation energy distribution exhibits a distinct preference for substitutional sites (orange curve), with a mean value for the segregating sites (i.e., $E_{seg}\leq 0\,\text{eV}$) of $-0.46\,\text{eV}$, indicating strong segregation to these sites. The interstitial segregation energy distribution (blue curve) has a higher mean segregation energy of $-0.22\,\text{eV}$, signifying a weaker attraction towards interstitial positions. Another difference that can be observed between the two spectra is that the majority (approx. 74\,\%) of interstitial sites is energetically unfavorable for P segregation (i.e., exhibiting so-called anti-segregation behavior, $\Delta E_{seg}^k(\text{P})>0\,\text{eV}$), while all substitutional sites are segregating. However, although P shows a stronger mean segregation tendency to substitutional sites, the number of interstitial sites, as shown in the inset plot, with similar segregation energies is approximately two times higher. This clearly demonstrates the importance of interstitial P segregation in $\alpha-$Fe. 

It is evident that the segregation behavior of Ni is markedly distinct. The substitutional segregation energies exhibit a mean value of approximately $-0.1\,\text{eV}$, indicating a somewhat weaker driving force for segregation than what is found for P. Notably, all interstitial sites demonstrate anti-segregating behavior with positive segregation energies. It can thus be concluded that Ni atoms will segregate exclusively to substitutional sites. 
The segregation energy distribution of H exhibits a contrasting behavior. The segregation energy to mean interstitial sites is $-0.16\,\text{eV}$, whereas all substitutional segregation energies are positive. Another difference emerges when comparing the H bulk formation energies. Supplementary Material Fig.~S1 shows that the tetrahedral position is only slightly more stable ($\approx 0.05\,\text{eV}$ for the used EAM potential) than the octahedral interstitial position. Since a typical unit of thermal energy at room temperature is $k_BT \approx 0.026\,\text{eV}$, these results suggest that octahedral bulk positions can serve as potential starting points for $\text{H}$ segregation. The cumulative segregation spectrum is shown in Supplementary Material Fig.~S4. However, the change is small, shifting the mean segregation energy from $-0.16\,\text{eV}$ to $-0.17\,\text{eV}$. 
Also noteworthy are the multiple peaks in the distribution of the interstitial segregation spectra of P and H. Similar work investigating the segregation behavior of H in fcc Ni~\cite{ding2024hydrogen} and fcc Pd~\cite{wagih2023spectrum}, respectively, concluded that these belong to different groups of interstitial sites with characteristic fingerprints divided into sites located in the core and at the surface of the GB, respectively.
 
A comprehensive overview of literature data of DFT calculated segregation energies to different coincidence site lattice GBs for these solutes is provided in graphical and tabular form in Supplementary Material Fig.~S2 and Table~S1. 
From this, it can be seen that a general consensus in existing literature exists that classical substitutional elements in bulk, such as Ni, Cr, Mo, etc, also segregate to substitutional sites at GBs. The same consensus exists for small interstitial atoms such as C and H and their segregation to interstitial sites. These trends are perfectly reproduced by our results showing that H does not segregate to substitutional sites and Ni does not segregate to interstitial sites. 
This indicates the reliability of the here proposed approach using Voronoi tesselation with MD and ML for obtaining the full interstitial segregation spectrum.

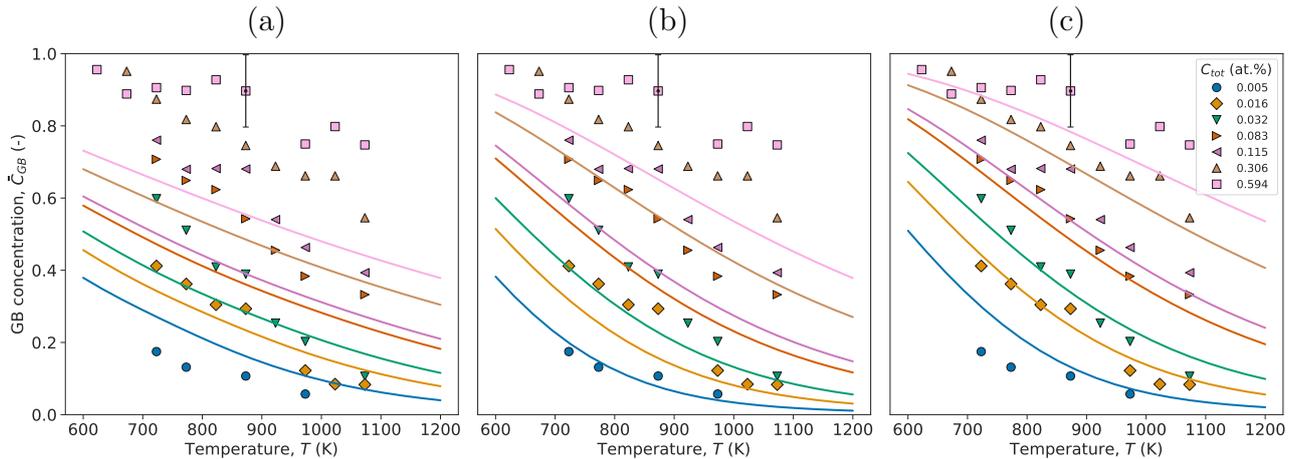
\begin{figure}[!ht]
    \centering
    \begin{picture}(0,0)
        \put(-150,0){(a)}
        \put(0,0){(b)}
        \put(150,0){(c)}
    \end{picture}\smallskip\\
    \includegraphics[width=1\textwidth]{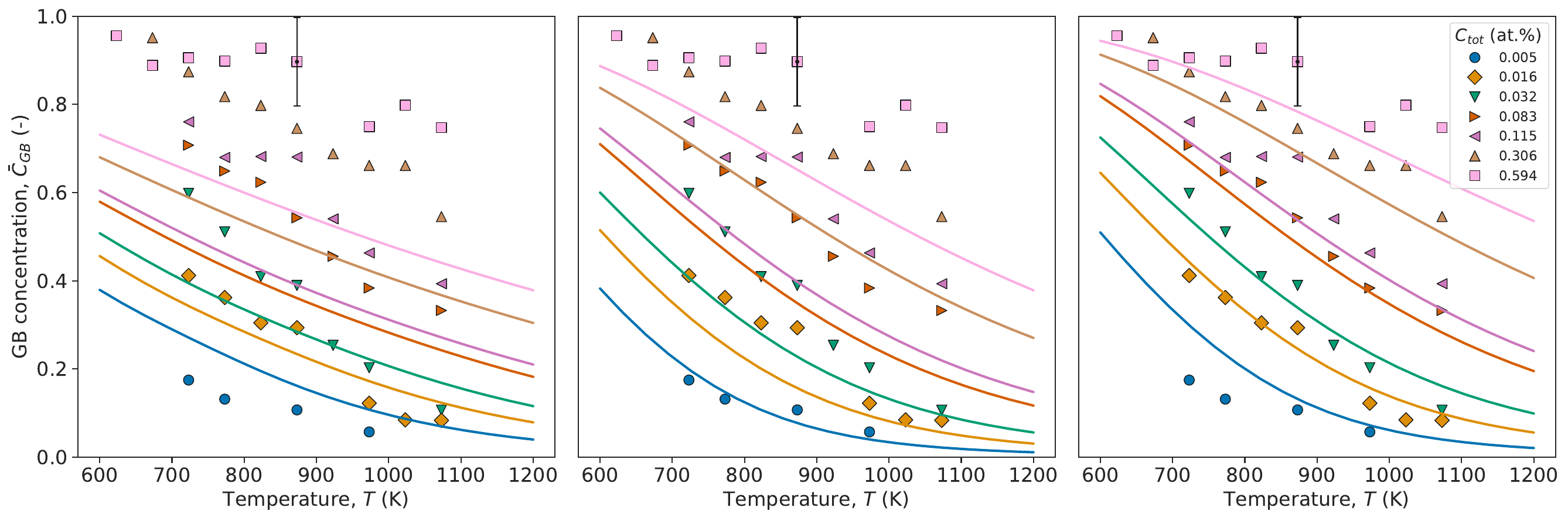}
    \caption{Comparison of experimental (symbols) and calculated GB concentrations of P ($\bar{C}_{GB}$) for different P concentrations ($C_{tot}$). a) Results using only the explicitly calculated EAM substitutional segregation energies of 100 representative sites, b) isotherms calculated for the entire substitutional segregation spectrum ($\approx 1.08\times10^5$ sites) obtained via ML, c) concentrations with both substitutional and interstitial segregation spectra included. The error bar represents the standard deviation of the measured GB concentrations, which is the same for all experimental data points taken from Ref.~\cite{grabke1987effects}.}
    \label{fig:p_enrichment_subplot}
\end{figure}

\newpage
With the P segregation spectra in hand, we proceed with the evaluation of the average GB solute concentrations ($\bar{C}_{GB}$) using Eq.~\ref{eqn:Conc_multisite} and compare these with the experimental findings of Erhart and Grabke~\cite{erhart1981equilibrium}. Analogous segregation isotherms for Fe-Ni and Fe-H systems can be found in Supplementary Material, Fig.~S5.
In general, the calculated GB concentrations increase with increasing P content and decrease with increasing temperature, in agreement with the experiments, as can be seen in Fig.~\ref{fig:p_enrichment_subplot}. However, the quantitative agreement with the experimental measurements clearly depends on the segregation spectra taken into account.

Starting from Fig.~\ref{fig:p_enrichment_subplot}a, where the isotherms are derived from the distribution of segregation energies explicitly calculated for 100 sites, obtained as the $k$-means representative cluster centers for the substitutional GB sites, the results obviously do not provide close agreement with the experimental data. Especially for the lowest and highest total P concentrations ($C_{tot}$), there is a significant disagreement with the experimental data. A quantitatively similar result was also obtained in \cite{reichmann2024grain}, where a spectrum with even more explicitly calculated segregation energies was used. The authors tried to resolve the discrepancy w.r.t. experiment by including different contributions in their atomistic studies, e.g., vibrational entropy and also solute-solute interactions. However, these led to a decrease in the GB concentrations and hence increased the disagreement with the AES data. 

In Fig.~\ref{fig:p_enrichment_subplot}b, we plot the GB concentrations calculated using the complete substitutional segregation spectrum, which includes $\approx 1.08\times10^5$ GB sites evaluated using our ML model. This results in a general spreading of the isotherms, improving the agreement with the AES data not only for high total P concentrations but also for low ones. Additionally, incorporating all GB sites increases the curvature of the isotherms, yielding even better fits with the experimental data. This highlights the crucial importance of the correct segregation energy spectrum, leaving the used interatomic potential or structural details of the actual polycrystalline model to be only minor contributions.

Figure~\ref{fig:p_enrichment_subplot}c presents the cumulative effect of interstitial and substitutional segregation states. Although the calculated GB concentrations in Fig.~\ref{fig:p_enrichment_subplot}b are in good agreement for low P concentrations up to approx. $0.016\,\text{at.\%}$, we see an increasing deviation for higher P contents. 
Including the segregation to interstitial sites (Fig.~\ref{fig:p_enrichment_subplot}c) shifts the isotherms to higher concentrations, matching the experimental data points almost perfectly for all considered P contents. This underlines the importance of considering the interstitial segregation spectrum, as the number of interstitial sites with similar energies to those of substitutional segregation is significantly higher. In Supplementary Material Fig.~S3, the relative contributions of substitutional and interstitial sites to GB enrichment are shown. This highlights how the contribution of substitutional and interstitial sites decreases and increases, respectively, with temperature. The inflection point on these curves marks the border between substitutional- and interstitial-dominated P segregation. It occurs at $\approx 900\,\text{K}$, which is very similar to the $700\,\text{K}$ reported by Lejček \etal in Ref.~\cite{lejvcek2016interstitial}.

Note that interstitial sites modify the average GB concentration not only through the different segregation spectrum but also via the denominator in Eq.~\ref{eqn:Conc_multisite}. For Fe-P, this has a general tendency to flatten the curve of $\bar{C}_{GB}(P)$ as a function of $T$. Together with the complete segregation spectra (both substitutional and interstitial), it brings the computational result into good agreement with the experimental data. Finally, solute-solute interactions between the different site types may influence the agreement with experiments, particularly at higher solute concentrations. However, due to computational complexity, these interactions are not considered here. Investigating their potential impact on the segregation isotherms is left to future work.

\newpage
In conclusion, we analyzed the substitutional and interstitial segregation spectra of P in $\alpha$-Fe using a polycrystalline GB model. Interstitial sites were selected based on Voronoi polyhedral construction, resulting in approximately $1.19\times10^6$ potential segregation sites in the GB regions, in addition to $1.08\times10^5$ substitutional (lattice) sites. The complete segregation energy spectrum was achieved with the help of a machine learning model based on explicit calculations of segregation energies of 100/300 representative substitutional/interstitial segregation sites using classical interatomic potentials. For comparison, we also examined the segregation behavior of Ni and H, elements for which the existing literature predominantly assumes segregation to substitutional and interstitial sites, respectively. The main findings of this study are as follows:
\begin{itemize}
    \item Our approach is able to reproduce the expected segregation behavior of Ni and H. The results indicate a preference for substitutional sites for Ni and interstitial sites for H. These findings are consistent with the existing literature and validate the methodology employed in this study. For the solute P, the spectra of the segregation energies are markedly different, showing a preference for substitutional sites.
    \item The mean negative segregation energy of P at substitutional GB sites ($-0.46\,\text{eV}$) is lower than that at interstitial GB sites ($-0.22\,\text{eV}$), indicating a preference for substitutional segregation. Furthermore, most interstitial sites are energetically unfavorable for P segregation (i.e., they are anti-segregating), while all examined substitutional sites are energetically favorable for segregation. However, due to the larger number of available interstitial sites, the total number of occupied interstitial sites is significantly higher than the number of substitutional sites. This helps explain discrepancies between experimental and theoretical studies of P segregation in $\alpha$-Fe. Previous computational studies, which were generally limited to a small subset of $\Sigma$ GBs, may have underestimated the influence of interstitial sites by neglecting the full spectrum of segregation energies. Additionally, the insufficient resolution of experimental measurements, which cannot easily differentiate between interstitial and substitutional segregation, has contributed to these discrepancies.
    \item By considering both interstitial and substitutional \emph{complete} segregation spectra, we achieved qualitative and quantitative agreement between experimentally measured GB enrichment of P and the calculated values across a wide temperature and concentration range. This emphasizes the crucial role of interstitial P segregation in accurately modeling GB enrichment.
    \item It is found that the accuracy of the applied machine learning method using SOAP features based on the undecorated GB model depends strongly on the amount of structural relaxation induced by the segregating species. Thus, the accuracy for substitutional solute segregation decreases in the order of Ni, P, H and for interstitial solute segregation in the order of H, P, Ni, represented by the size mismatch between solute and GB site.
\end{itemize}

\section*{Acknowledgment}
The financial support by the Austrian Federal Ministry for Labour and Economy, and the Christian Doppler Research Association is gratefully acknowledged.
Part of this research was funded by the Austrian Science Fund (FWF) \mbox{[P 34179-N]}. 
This work was carried out (partially) using supercomputer resources provided under the EU-JA Broader Approach collaboration in the Computational Simulation Centre of the International Fusion Energy Research Centre (IFERC-CSC).
Calculations were partly performed using supercomputer resources provided by the Vienna Scientific Cluster (VSC).


 \bibliographystyle{elsarticle-num} 
 \bibliography{cas-refs}

\end{document}